\documentclass[10pt,a4paper]{article}

\usepackage[utf8]{inputenc}

\usepackage[utf8]{inputenc}
\usepackage[T1]{fontenc}

\usepackage{cite}
\usepackage{hyperref}
\usepackage{url}    

\usepackage{amsmath,amssymb,amsfonts}
\usepackage{algorithm}
\usepackage{algorithmicx}
\usepackage{algpseudocode}
\usepackage[caption=false]{subfig}
\usepackage{graphicx}
\usepackage{textcomp}
\usepackage{xspace}
\usepackage{xcolor}
\usepackage{tikz}
\usepackage{hhline}
\usepackage{pgfplots}
\pgfplotsset{compat=newest,compat/show suggested version=false}

\usepackage{cancel} 
\usepackage{wasysym} 
\usepackage{multirow} 

\usepackage{mathtools}
\DeclarePairedDelimiter\ceil{\lceil}{\rceil}
\DeclarePairedDelimiter\floor{\lfloor}{\rfloor}

\newcommand{\tree}{\ensuremath{\text{$b$-tree}}\xspace}
\newcommand{\build}{\textsc{BuildTree}\xspace}

\newcommand{\find}{\textsc{FindNeighbors}\xspace}

\algdef{SE}[DOWHILE]{Do}{DoWhile}{\algorithmicdo}[1]{\algorithmicwhile\ #1}%
\algnewcommand{\algorithmicgoto}{\textbf{go to}}%
\algnewcommand{\Goto}[1]{\algorithmicgoto~\ref{#1}}%

\begin{document}

\title{Finding Neighbors in a Forest: A \textit{$b$-tree} for \\Smoothed Particle Hydrodynamics Simulations}

\author{
Aurélien Cavelan\\
\textit{University of Basel, Switzerland}\\
\textit{aurelien.cavelan@unibas.ch}
\and
Rubén M. Cabezón\\
\textit{University of Basel, Switzerland}\\
\textit{ruben.cabezon@unibas.ch}
\and
Jonas H. M. Korndorfer\\
\textit{University of Basel, Switzerland}\\
\textit{jonas.korndorfer@unibas.ch}
\and
Florina M. Ciorba\\
\textit{University of Basel, Switzerland}\\
\textit{florina.ciorba@unibas.ch}}


\maketitle

\begin{abstract}
Finding the exact close neighbors of each fluid element in mesh-free computational hydrodynamical methods, such as the Smoothed Particle Hydrodynamics (SPH),
often becomes a main bottleneck for scaling their performance beyond a few million fluid elements per computing node. 
Tree structures are particularly suitable for SPH simulation codes, which rely on finding the exact close neighbors of each fluid element (or SPH particle).
In this work we present a novel tree structure, named \textit{$b$-tree}, which features an adaptive branching factor to reduce the depth of the neighbor search.
Depending on the particle spatial distribution, finding neighbors using \tree has an asymptotic best case complexity of $O(n)$, as opposed to $O(n \log n)$ for other classical tree structures such as octrees and quadtrees.
We also present the proposed tree structure as well as the algorithms to build it and to find the exact close neighbors of all particles.
We assess the scalability of the proposed tree-based algorithms through an extensive set of performance experiments in a shared-memory system. 
Results show that \tree is up to $12\times$ faster for building the tree and up to $1.6\times$ faster for finding the exact neighbors of all particles when compared to its octree form.
Moreover, we apply \tree to a SPH code and show its usefulness over the existing octree implementation, where \tree is up to $5\times$ faster for finding the exact close neighbors compared to the legacy code.
\end{abstract}


\section{Introduction}
\label{sec:intro}

Hydrodynamical simulations rely, in their vast majority, on efficient algorithms for finding the exact close neighbors among the fluid elements. 
In the case of static meshes, the neighborhood is inherently fixed by the mesh geometry, and can be naively explored for finding neighbors using stencils. 
But, how can neighbors be found in unstructured meshes?
This is precisely the case of the Smoothed Particle Hydrodynamics (SPH) technique~\cite{gingold1977, lucy1977, springel2010, rosswog2015}. 
This method discretizes the fluid in a series of interpolating points (named SPH particles or particles, hereafter) that are distributed following the actual density profile of the simulated fluid. 
Then the physical properties of each particle are obtained with a weighted radial interpolation over closely neighboring particles. 
The weight of this interpolation has compact support, and its radius is named the \textit{smoothing length}. 
This means that in non-homogeneous, highly dynamic systems --such as those found in Computational Fluid Dynamics (CFD) and Astrophysics-- the particle distribution can be geometrically very distorted. 
In this case, finding close neighbors in a computationally-efficient way is a non-trivial problem. 


The most common approach for finding neighbors in SPH is to use a tree structure that recursively divide the spatial computational domain in sub-cells\footnote{In the rest of this paper, we refer indifferently to a tree node as a cell and to its children as a sub-cells.}, until leaves are empty, or contain a small bucket of particles.
Once the tree is built, it can be walked to find the exact close neighbors of each particle, discarding whole branches when their parent cells are too far from the current particle, thereby decreasing the \textit{search time}.
This is particularly relevant in Astrophysical simulations, where the tree also stores information about the multipolar expansion of the gravitational field and is used to efficiently evaluate the gravitational force that the particles experience~\cite{hernquist1989}.

In this work, we introduce a novel tree algorithm for the \textit{exact} close neighbor search, named \textit{\tree}. The proposed tree aims at drastically decreasing the \textit{search time} by building a very shallow tree. 
This is achieved by choosing a very high branching factor, i.e. allowing nodes to have many children, effectively building a \textit{broad-tree} instead of a deep-tree. 

\tree uses an adaptive branching factor to prevent the number of sub-cells from increasing excessively within a single cell of the tree.
This is done by enforcing a limit both, on the maximum number of particles per cell (i.e., the bucket size) and on the number of empty sub-cells allowed.

More specifically, \tree recursively divides the spatial computational domain into smaller, \textit{equal-size sub-cells}. The resulting sub-cells are then mapped onto a regular grid structure for easy access: grid structures with equal-size cells render an $O(1)$ access time, i.e., it is possible to find the sub-cell that contains a given particle with known coordinates in constant time\footnote{In theory, it is possible to map any simulation domain onto a large enough grid. In practice, the number of grid cells required would quickly render this design impractical. Specifically, this approach relies on grids with equal-size cells: the cell size (and, therefore, the number of cells in the grid) needs to accommodate the most dense part of the computational domain. To guarantee that any cell in the grid contains at most one particle, the cell size must be smaller than the shortest distance between any two particles. For very heterogeneous particle distributions, this approach would require a prohibitive amount memory to be a viable solution, hence the adaptive branching factor proposed herein.} (see Section~\ref{sec:new_tree}).

The resulting tree has interesting properties. 
In particular when the set of $n$ particles has a relatively uniform distribution across the computational domain, the depth of \tree is $1$ and requires $O(n)$ steps to be built and $O(n)$ steps to find the neighbors of $n$ particles (i.e., a constant number of steps per particle), as opposed to $O(n \log n)$ in the best case with a standard octree implementation.
A classical octree requires at most $O(n d \log n)$ steps to recursively build the tree, where $d$ is the depth of the tree, and at most $O(n d)$ steps to find the neighbors (i.e., $O(d)$ per particle), where $d$ is typically close to $\log n$.

In this work, we experiment with both, a uniformly distributed 3D particle dataset as well as a non-uniformly 3D particle dataset, to demonstrate the \tree properties and performance benefits. 
The code that builds the tree and finds exact neighbors is provided as a standalone code, with both C++ and Fortran interfaces, while the experiments are provided as a single reproducible package.
Furthermore, \tree has been integrated into an astrophysical SPH code, SPHYNX~\cite{cabezon2017}.
However, the applicability of the \tree code is neither limited to astrophysical applications, nor SPH codes.

The remainder of the work is structured as follows.
In Section~\ref{sec:related_work} we review the different tree algorithms that can be found in the SPH literature. 
We introduce the proposed \tree algorithm in Section~\ref{sec:new_tree}. 
We evaluate its performance against the classical octree algorithm in Section~\ref{sec:experiments}. 
We conclude the work in Section~\ref{sec:conclusions} and outline future work directions.

\section{Related Work}
\label{sec:related_work}

A vast amount of literature exists on tree-based algorithms.
A detailed review can be found in the work of Curtin et al.~\cite{curtin2015} (and references therein).
In this section, we review three methods in the field of CFD and Computational Astrophysics, with a particular focus on SPH simulation codes.

\indent\textbf{Octrees}~\cite{jackins1980} have widely been used in sinergy with SPH codes. 
With octrees, the computational domain is recursively halved in each dimension, until there is only a single particle per leaf or none at all. 
Octrees are a 3D generalization of quadtrees that are typically applied to 2D computational domains~\cite{finkel1974}. 
Hernquist~\&~Katz~\cite{hernquist1989} first proposed the usage of a hierarchical tree structure as an efficient, dynamical, and fully-Lagrangian method to evaluate gravitational forces. 
This method was based on the Barnes-Hut algorithm~\cite{barnes1986} that uses an octree to evaluate gravitational forces within a multipolar approximation. 
Octrees are used in GADGET2~\cite{springel2005}, ChaNGa~\cite{menon2015}, SPH-flow~\cite{oger2016}, and SPHYNX~\cite{cabezon2017}, among many other SPH codes.
 
\indent\textbf{\textit{kd}-trees}~\cite{bentley1975} have been proposed to avoid the exponential dependence of quadtrees and octrees to the spatial dimension. 
In \textit{kd}-trees, each node
has only two children, but every division is always aligned to one of the dimension axes. 
\textit{kd}-trees are employed by ~\cite{wadsley2017,price2017,emmanuel,stadel} in GASOLINE2, PKDGRAV2, PHANTOM, and other general purpose SPH codes. 
 
\indent\textbf{Ball trees}~\cite{omohundro1989} are
binary trees in which each node has an associated hyper-sphere that it is the smallest volume that contains the hyper-spheres of its children. 
Unlike \textit{kd}-trees, the node regions can intersect and do not require the partitioning of the entire computational domain. 
Ball trees are seldomly used by SPH codes. 

\indent\textbf{\tree vs. octree}:
While octrees have branching factors of $2^3$ on average, i.e. halving the spatial domain along each dimension in 3D distributions, \tree may have branching factors up to $n$, depending on the particle spatial distribution. 
The number of children at each node in \tree depends on the branching factor ($b$) and on the bucket size ($s$) that determines the maximum number of particles in each leaf. 
$s$ is a user-defined parameter, while $b$ is computed while building the tree in order to adapt to the particle distribution.
In the best case, for perfectly uniform particle distributions, \tree has $O(1)$ complexity for finding the exact nearest neighbors of a particle, while octrees have $(\log n)$ complexity at best.
In the worst case, when the particle distribution is found to be highly non-uniform, \tree collapses to an octree and $b$ is automatically set to $2$ and the complexity for finding the exact nearest neighbors of a particle is $O(d)$ for both.
Table~\ref{table.symbols} shows the different parameters used in this paper.

\indent\textbf{\tree vs. stratified trees}: \tree has a high branching factor, which is a characteristic shared with stratified trees, such as the van Emde Boas trees~\cite{vanEmde75}. 
Stratified trees have been used for finding the approximate nearest neighbor~\cite{amir1999,cary2001}.
with lower complexity, namely $O(\log\log n)$, than that of octrees, namely $O(\log d)$ for imbalanced trees and $O(\log n)$ for balanced trees.
However, hydrodynamical simulations require finding the exact neighbors of each particle, and \tree provide exactly this, at the cost of sacrificing part of the asymptotic complexity of stratified trees, yet still being more efficient than octrees.

\begin{table}
	\caption{Notation used in this work}
	\label{table.symbols}
	\begin{center}
		\begin{tabular}{ c | l}
			\hline
			\multicolumn{2}{ c }{\textbf{Input Parameters}} \\
			\hline
			Name & Description \\
			\hline
			$k$ & Number of spatial dimensions \\
			$n$ & Total number of particles \\
			\hline
			\multicolumn{2}{ c }{\textbf{Tree Parameters}} \\
			\hline
			$s$ & Bucket size, i.e. maximum number of particles in a leaf \\
			$\alpha$ & Fraction of the bucket size $s$ \\ 
			$\beta$ & Max. ratio of cells with less than $\alpha S$ particles \\
			\hline
			\multicolumn{2}{ c }{\textbf{Other Variables}} \\
			\hline
			$d$ & Tree depth \\
			$b$ & Branching factor per spatial dimension (at a given node)\\
			$r$ & Current ratio of cells with less than $\alpha S$ particles \\
			$n_i$ & Number of particles in cell $i$ \\
		\end{tabular}
	\end{center}
\end{table}


\section{\tree}
\label{sec:new_tree}

In this section, we introduce the proposed \tree structure and present the algorithms for building the tree in Section~\ref{sec.building} and for finding the exact close neighbors of SPH particles in Section~\ref{sec.findneighbors}.

\subsection{Tree Building}
\label{sec.building}

Algorithm~\ref{fig.algo.build} describes the \build process. 
The algorithm recursively distributes a given set of $n$ particles into smaller, equal-size $b^k$ sub-cells, where $k$ is the number of spatial dimensions and $b$ is the branching factor, i.e. the number of children in each dimension for the current node (see Table~\ref{table.symbols}).


\begin{algorithm}
\caption{BuildTree}
\label{fig.algo.build}
\begin{algorithmic}
\Procedure{BuildTree}{$n$, $i$}
		\State $redistribution \gets true$
		\While{$redistribution$ is $true$}
		\State Compute branching factor $b$ (Equation~\ref{eq.branchingfactor})
		\State Create $b^k$ cells and distribute the particles:
			\For{each particle $p$ in cell $i$}
				\State Compute its sub-cell coordinates (Equation~\ref{eq.rounding})
				\State Compute its sub-cell id $j$ (Equation~\ref{eq.cell-index})
				\State Add $p$ to its corresponding sub-cell $j$
				\State Update $n_j$
			\EndFor
		\State Compute the distribution ratio $r$ (Equation~\ref{eq.ratio})
		\If{$R < \beta$}
			\State $b = \frac{b}{2}$
		\Else
			\State $redistribution \gets false$
		\EndIf
		\EndWhile
		\For{each sub-cell $j$}
			\If{$n_j > s$}
				\State \Call{BuildTree}{$n_j$, $j$}
			\EndIf
 	  	\EndFor
\EndProcedure
\end{algorithmic}
\end{algorithm}


\textbf{Step 1}: Given a set of $n$ particles in the current cell $i$, the \build algorithm computes the branching factor $b$ and create $b^k$ empty equal-size sub-cells.
To compute the branching factor $b$, we initially assume that particles are uniformly distributed in the spatial computational domain. 
Therefore, given an upper limit on the number of particles per bucket $s$, we want to find $b$ such that $\frac{n}{b^k} \leq s$.
Solving for $b$, and rounding to the nearest higher integer value, we obtain:
\begin{align}
b = \ceil*{\sqrt[k]{\frac{n}{s}}} \ .
\label{eq.branchingfactor}
\end{align}


\textbf{Step 2}: 
The particles are distributed into the newly created cells, based on the particles coordinates.
The sub-cells are mapped onto a $k$-dimensional grid, i.e. each sub-cell has a cut of the current computational domain.
Each sub-cell therefore has its own set of $k$D-coordinates within the grid.
For each particle with coordinates $x_1,\cdots,x_k$, we compute its corresponding sub-cell coordinates $x_1',\cdots,x_k'$ by:
(1) normalizing the coordinates of the particle with respect to the current cell's sub-domain;
(2) subsequently multiplying the normalized coordinates by the number of sub-cells in each dimension, i.e., $b$; and 
(3) rounding the resulting coordinates to the nearest lower integer value as follows:
\begin{align}
x_l' = \floor*{\frac{x_l-x_{l,min}}{x_{l,max}-x_{l,min}} \cdot b},
\label{eq.rounding}
\end{align}
where $x_{l,min}$ and $x_{l,max}$ are the bounds of the domain in the $l$ dimension (otherwise known as bounding box).

For example, consider a particle with the following 3D coordinates $x_1=3.6$, $x_2=4.2$, and $x_3=0.6$. You cannot change the nomenclature to x,y,z when we are using a generalized coordinate system that is valid to all coordinate sets.
In this example, $b$ has been set to $10$. 
Suppose that this particle is assigned to the sub-domain box defined by $x_{1,min}=x_{2,min}=x_{3,min}=0$ and $x_{1,max}=x_{2,max}=x_{3,max}=5$. 
According to Eq.~\ref{eq.rounding}, we obtain the coordinates of the cell that contains the particular particle as follows:

$$x_1' = \floor*{\frac{3.6}{5} \cdot 10} = \floor*{7.2} = 7,$$
$$x_2' = \floor*{\frac{4.2}{5} \cdot 10} = \floor*{8.4} = 8,$$
$$x_3' = \floor*{\frac{0.6}{5} \cdot 10} = \floor*{1.2} = 1.$$


For each particle in the domain, assuming that cells are stored in a 1D array and numbered from $0$ to $W^d-1$, we can compute the corresponding sub-cell id $j$ as follows:
\begin{align}
j = x_1' + x_2' W + x_3' W^2 \ .
\label{eq.cell-index}
\end{align}
In our example, the index of the corresponding bucket within the current node would be $j = 7 + 8 \cdot 10 + 1 \cdot 10^{2} = 187$ (out of $b^3 = 1000$ sub-cells).

\textbf{Step 3}: Finally, to prevent the tree breadth from exploding if the initial value of $b$ was too high, we introduce two parameters, namely $\alpha$ and $\beta$ (see Table~\ref{table.symbols}), to control the number of sub-cells per node. 
Specifically, we first compute the distribution ratio $r$, which measures the ratio of sub-cells that contain less than $\alpha s$ particles (with $\alpha \sim 0.5$) over the total number of sub-cells $b^d$ within the current node:
\begin{align}
r = \frac{\sum_{j=0}^{W^d-1} n_j \leq \alpha S}{W^d} \ ,
\label{eq.ratio}
\end{align}
where $n_j$ is the number of particles that have been assigned to sub-cell $j$.
If the resulting distribution ratio $r$ is larger than $\beta$ (with $\beta\sim 0.5$), it means that too many cells with too few particles have been created within the current node. Therefore, we divide the branching factor $b$ by $2$, hence reducing the number of cells by $2$ in each dimension, or by a factor of $2^{k}$ in total.

We recompute steps $1$, $2$ and $3$ until this criterion is met, i.e., when $r < \beta$. In the worst case, the algorithms stops at $b=2$, which corresponds to an octree structure.

The particle redistribution takes at most $O(\log b)$ steps, which is proportional to $O(\log n)$ due to $\log b = \log \left(\frac{n}{s}\right)^{\frac{1}{d}} = \frac{1}{d} \log \frac{n}{s}$. 

\subsection{Finding Neighbors}
\label{sec.findneighbors}

Algorithm~\ref{fig.algo.find} describes the \find algorithm.
In SPH simulations, each particle is characterized by a \textit{smoothing length}, denoted $h$. 
Finding the exact close neighbors of a given particle reduces to finding all particles that are within the $2h$ radius\footnote{$2h$ is the standard neighborhood radius of many interpolating functions (or kernels) in SPH. Although it is possible to use a kernel that employs a larger radius, the radii are always proportional to the smoothing length $h$. The \tree approach proposed herein is directly applicable to other SPH kernels with larger such radii.}.

To find the exact close neighbors of a given particle, the algorithm needs to walk the tree by discarding cells that are not within the $2h$ radius, and by visiting cells that are within the $2h$ radius.

More specifically, given a particle $p$, its radius $2h_p$ and a starting cell $i$, the algorithm needs to identify the sub-cells that are within the $2h_p$ radius of $p$. Rather than checking if every sub-cell is within the $2h_p$ radius, the algorithm directly computes the range of sub-cells coordinates to visit in each dimension $l$, denoted by $r_l = [r_{l,min},r_{l,max}]$.

\begin{figure}
	\begin{center}
		\begin{tikzpicture}
			\fill [blue,fill opacity=.3] (2,3) rectangle (5,5);
			\draw [step=1,black,thick] (0,0) grid (6,6);
			\draw (3.5,3.9) circle (0.8cm);
			\draw[black] plot [only marks, mark=*, domain=0:6, samples=200, mark size=0.8] (\x,rand*3+3);
			\draw[->,thick,red] (3.5,3.9)--(3.5,4.7) node[midway,left]{$2h$};
			\draw[red] plot [only marks, mark=*, mark size=0.8] (3.5,3.9);
			\draw[blue] (2,3.2) node[left,fill=black!20]{$r_{x,min}=2$};
			\draw[blue] (5,5) node[above,fill=black!20]{$r_{x,max}=5$};
			\draw[blue] (2,3) node[below,fill=black!20]{$r_{y,min}=3$};
			\draw[blue] (5,4.8) node[right,fill=black!20]{$r_{y,max}=5$};
		\end{tikzpicture}
	\end{center}
	\caption{Range of sub-cells to visit (in blue) within the current node in order to find all the close neighbors of the particle highlighted in red, i.e. all particles that are within its $2h$ radius.}
	\label{fig.findneighbors}
\end{figure}
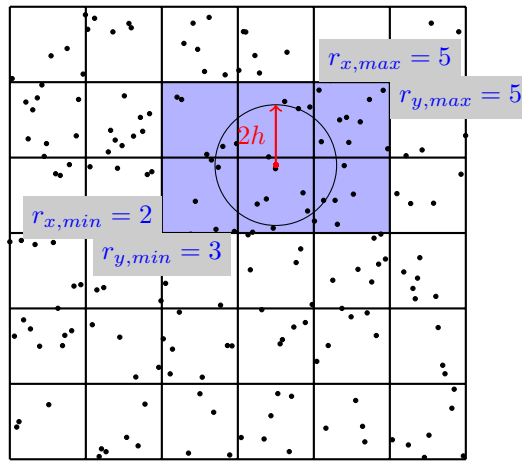

All close neighbors of $p$ are within the range $[x_l-2h_p,x_l+2h_p]$, with $l \in [1,k]$, and where $x_1,\hdots,x_k$ are the coordinates of the current particle $p$. Based on Equation~\ref{eq.rounding}, we write:
\begin{align}
r_{l,min} &= \floor*{\frac{(x_l-2h_p)-x_{l,min}}{x_{l,max}-x_{l,min}} \cdot b} \\
r_{l,max} &= \floor*{\frac{(x_l+2h_p)-x_{l,min}}{x_{l,max}-x_{l,min}} \cdot b}.
\label{eq.range}
\end{align}
Figure~\ref{fig.findneighbors} illustrates the process with a 2D example, showing which cells need to be visited in order to find the neighbors of the particle highlighted in red.

Then, for every tuple of sub-cell coordinates within the computed range, we retrieve the corresponding sub-cell id, $j$, using Equation~\ref{eq.cell-index}. When the algorithm reaches a cell $j$ that contains $n_j \leq s$ particles, then the cell is a leaf and the algorithm checks all the particles that are stored inside the cell individually. For every particle $q$ in cell $j$, we compute the Cartesian distance between $p$ and $q$, and if the distance is less than $2h_p$ we add particle $q$ to the list of neighbors of particle $p$, denoted by NeighborsOf($p$).

\begin{algorithm}
\caption{FindNeighbors}
\label{fig.algo.find}
\begin{algorithmic}
\Procedure{FindNeighbors}{$p$, $h_p$, $i$}
	\State Compute range $r_l = [r_{l,min},r_{l,max}]$, $l \in [1,k]$  (Eq.~\ref{eq.range})
	\For{each cell $(x_1, x_2, \hdots, x_k) \in r_1 \times r_2 \hdots \times r_k$}
		\State Compute sub-cell id $j$ (Eq.~\ref{eq.cell-index})
		\If{$n_j \leq s$}
			\For{each particle $q$ in cell $j$}
				\If{Distance($p$,$q$) $ \leq 2h_p$}
					\State Add $q$ to NeighborsOf($p$)
				\EndIf
			\EndFor
		\ElsIf{$n_j > s$}
			\State \Call{FindNeighbors}{$p$, $h_p$, $j$}
		\EndIf
	\EndFor
\EndProcedure
\end{algorithmic}
\end{algorithm}


\subsection{Complexity Analysis}

\textbf{Time complexity}: 
Overall, the number of steps required for building the tree and finding the neighbors depends on the depth of tree, denoted $d$.
The \build algorithm requires at most $O(n d \log n)$ steps for redistributing at most $n$ particles $\log n$ times, over $d$ levels, while the octree only requires $O(n d)$ steps, without the additional redistribution steps.
Both the \find algorithm and classical octree implementations requires at most $O(n d)$ steps in order to find the neighbors of $n$ particles, due to having to visit $d$ levels every time.

The main difference between our \tree and an octree is the depth of the tree, which depends on the particle spatial distribution.
In the worst case, \tree collapses to a quadtree (2D) or an octree (3D) with the same depth. 
Unlike $kd$-trees, the depth of a quadtree or an octree is not guaranteed to be $\log n$ in the worst case.
In fact, in the worst case there is only one particle per level and there are $d=n$ levels.
In the best case, when particles are distributed uniformly across the spatial computational domain, an octree will be balanced and have depth $d=\log n$, while \tree will be only one level deep and have depth $d=1$, and will require no redistribution.

Table~\ref{table.complexity} summarizes the time complexity for the proposed \tree, compared to a classical octree. $n$ is the number of particles and $d$ denotes the maximum depth of the tree.

\textbf{Memory complexity}:
In the worst case, \tree may create up to $n$ cells per level (with up to $d=n$ levels in the worst case), therefore the space required is at most $O(nd)$.
While \tree may have many more cells than an octree, only non-empty cells need to be stored in memory, and the impact remains small compared to the space required for storing the neighbors of every particle.


\begin{table}
\caption{Asymptotic complexity for tree building and finding neighbors: \tree and octree comparison}
\label{table.complexity}
\begin{tabular}{ c | c | c | c | c |}
\multirow{2}{*}{ }  & \multicolumn{2}{ c |}{\textbf{Tree Building}} & \multicolumn{2}{ c |}{\textbf{Finding Neighbors}} \\
 \cline{2-5}
& Worst Case & Best Case & Worst Case & Best Case \\
 \hline
\tree & $O(n d \log n)$ & $O(n)$ & $O(n d)$ & $O(n)$ \\ 
octree & $O(n d)$ & $O(n\log n)$ & $O(n d)$ & $O(n \log n)$ \\
\hline
\end{tabular}
\end{table}

\section{Experiments}
\label{sec:experiments}

In this section, we conducted experiments to assess the scalability of the proposed \tree in a shared-memory system as well as to evaluate the impact of the different parameters on its performance.
Moreover, \tree has been integrated into an astrophysical SPH code, SPHYNX~\cite{cabezon2017}, and the performance of the new tree-based algorithms are evaluated against the legacy code in Section~\ref{sec.sphynx}.


\subsection{Experimental Setup}

Building the tree and finding the neighbors are two operations that are typically performed within a single processing node, over a subset of the entire simulation domain, called sub-domain. 
The topic of domain partitioning across the computation nodes is beyond the scope of this work. 
In this work, we focus on the performance of the \build and \find algorithms for exploiting many-core parallelism.
Specifically, we perform experiments using a single Intel Xeon compute node, the details of which are presented in Table~\ref{table.node}. Execution times results are averaged out of 100 executions for every configuration to produce representative data.

\begin{table}[!htb]
\centering
\caption{Characteristics of the experimental platform}
\label{table.node}
\begin{tabular}{l|l}
\hline
\textbf{Parameter}           & \multicolumn{1}{l}{\textbf{Description}}                                         \\ \hline
Operating system    & \begin{tabular}[c]{@{}l@{}}CentOS Linux\\ release 7.2.1511\end{tabular} \\ \cline{2-2} 
Processor           & \begin{tabular}[c]{@{}l@{}}Intel Xeon\\ E5-2640 v4\end{tabular}         \\ \cline{2-2} 
Number of cores     & 20 (+20 with hyperthreading)                                            \\ \cline{2-2} 
Memory                 & 64 GB RAM                                                                  \\ \cline{2-2} 
Operating frequency & 2.4 – 3.4 GHz                                                           \\
\hline
\end{tabular}
\end{table}

We consider two SPH simulation test cases: (a) an Evrard collapse (EC) test~\cite{evrard1988} with $10^6$ particles, which studies the gravitational collapse of a gaseous cloud; 
EC is a common test to evaluate the correctness of the coupling of hydrodynamics and self-gravity and is an example of non-uniform particle distribution; and (b) a Square Patch (SP) test \cite{colagrossi2005} with $10^7$ particles; SP is a common test case in CFD to simulate highly distorted geometries and resistance to particle cumpling, and has a fairly uniform particle distribution.

\begin{table}[!htb]
\centering
\caption{Design of target experiments}
\label{fig.experiments}
\begin{tabular}{| c | c | c |}
\hline
\textbf{Parameter Name} & \multicolumn{2}{|c|}{\textbf{Default Value}} \\
\hline
\hline
Test case & EC & SP \\
\#Threads & $40$ & $40$ \\
Particle distribution & Non-uniform & Uniform \\
\#Dimensions ($d$) & 3 & 3 \\
\#Particles ($n$) & $10^6$ & $10^7$ \\
Target \#neighbors/particle & $100$ & $500$ \\
Bucket size ($s$) & $8$ & $8$ \\
$\alpha$ & $0.5$ & $0.5$ \\
$\beta$ & $0.5$ & $0.5$ \\
\hline
\end{tabular}
\end{table}

\subsection{Strong Scaling}

In this sub-section, we assess the speedup of the \tree-based \build and \find algorithms, with respect to their standard octree-based counterparts using the particle datasets of the EC and SP test cases.

\begin{figure*}[!htb]
	\centering
		\subfloat[Evrard Collapse]{\includegraphics[scale=0.8]{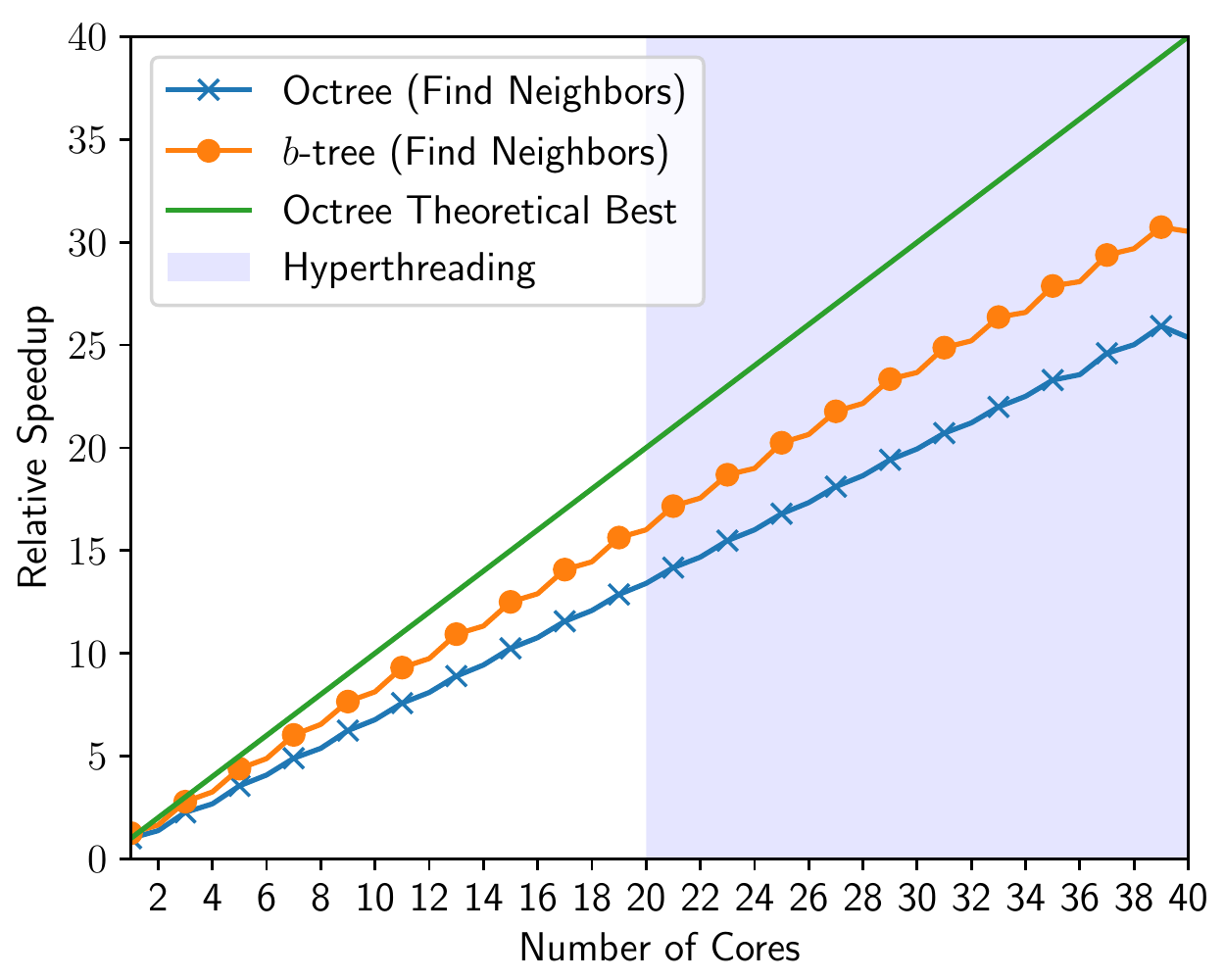}}\\
		\subfloat[Square Patch]{\includegraphics[scale=0.8]{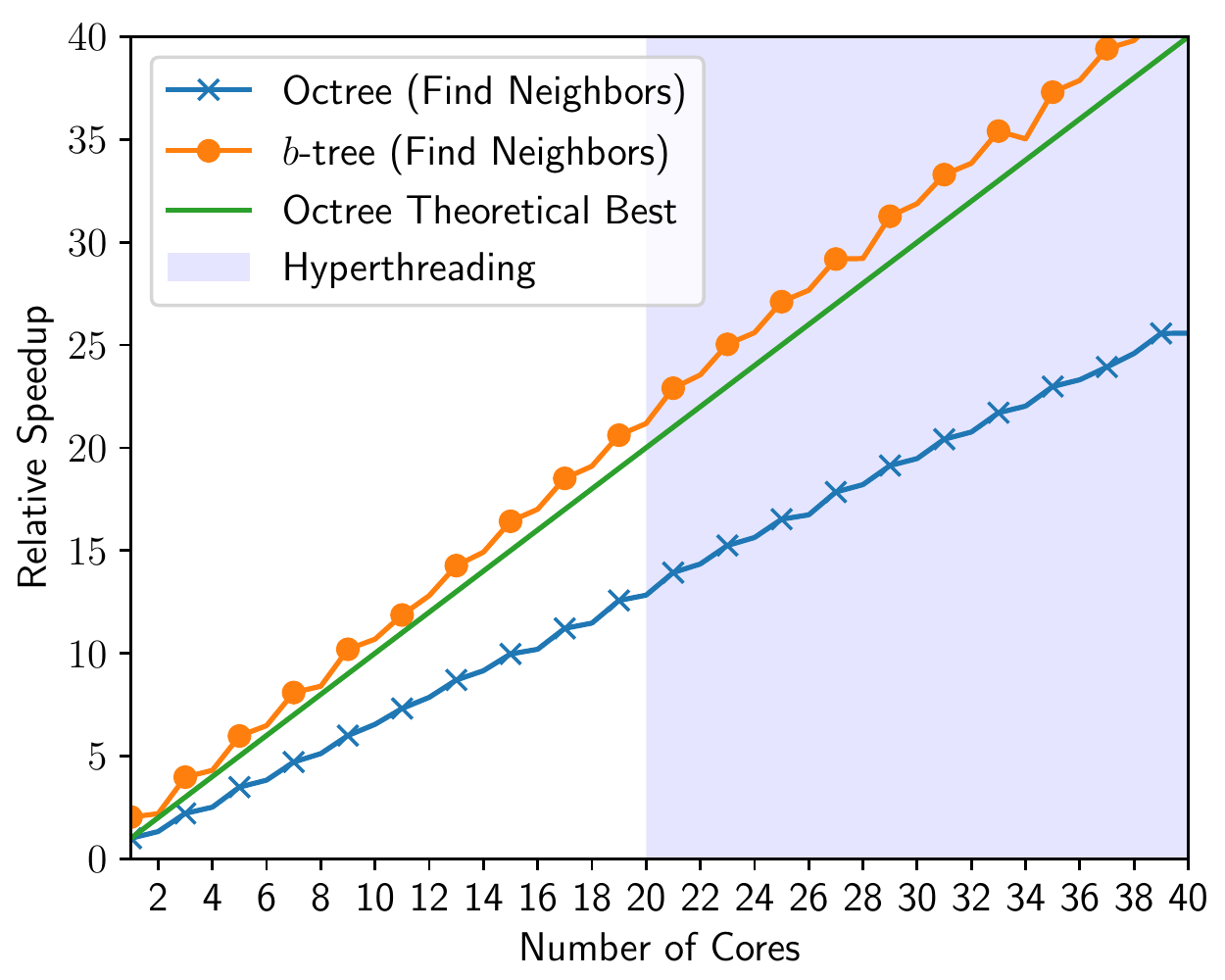}}
	\caption{\tree speedup normalized to the octree speedup for the EC test~(a) and the SP test~(b). The highlighted region denotes the experiments where hyper-threaded cores were employed in the experiments.}
	\label{fig.speedup}
\end{figure*}

The relative speedup for building the tree and finding the neighbors is computed with respect to the octree build and search time, respectively, with one core. 
Figure~\ref{fig.speedup}(a) and Figure~\ref{fig.speedup}(a) show the the relative speedup for EC and SP test cases, respectively. 
While both search algorithms scale very well, even when exploiting the hyper-threading available on the Intel Xeon, the proposed \find algorithm always yields better performance compared to the classical octree algorithm.
In addition, the proposed \build algorithm is always faster than its classical octree counterpart.





\subsection{Impact of the Bucket Size}

\begin{figure*}[!htb]
	\centering
		\subfloat[Evrard Collapse]{\includegraphics[scale=0.5]{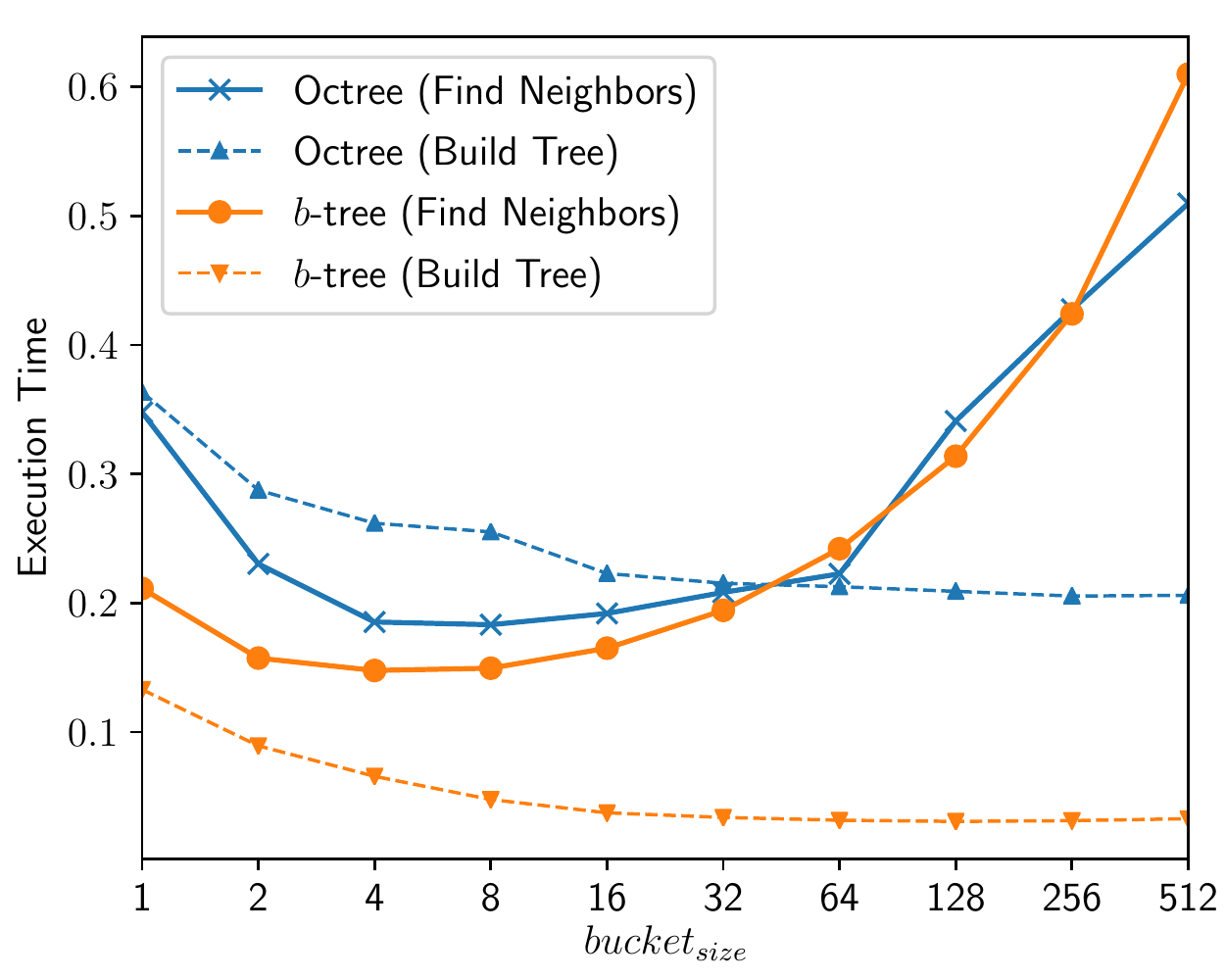}}
		\subfloat[Square Patch]{\includegraphics[scale=0.5]{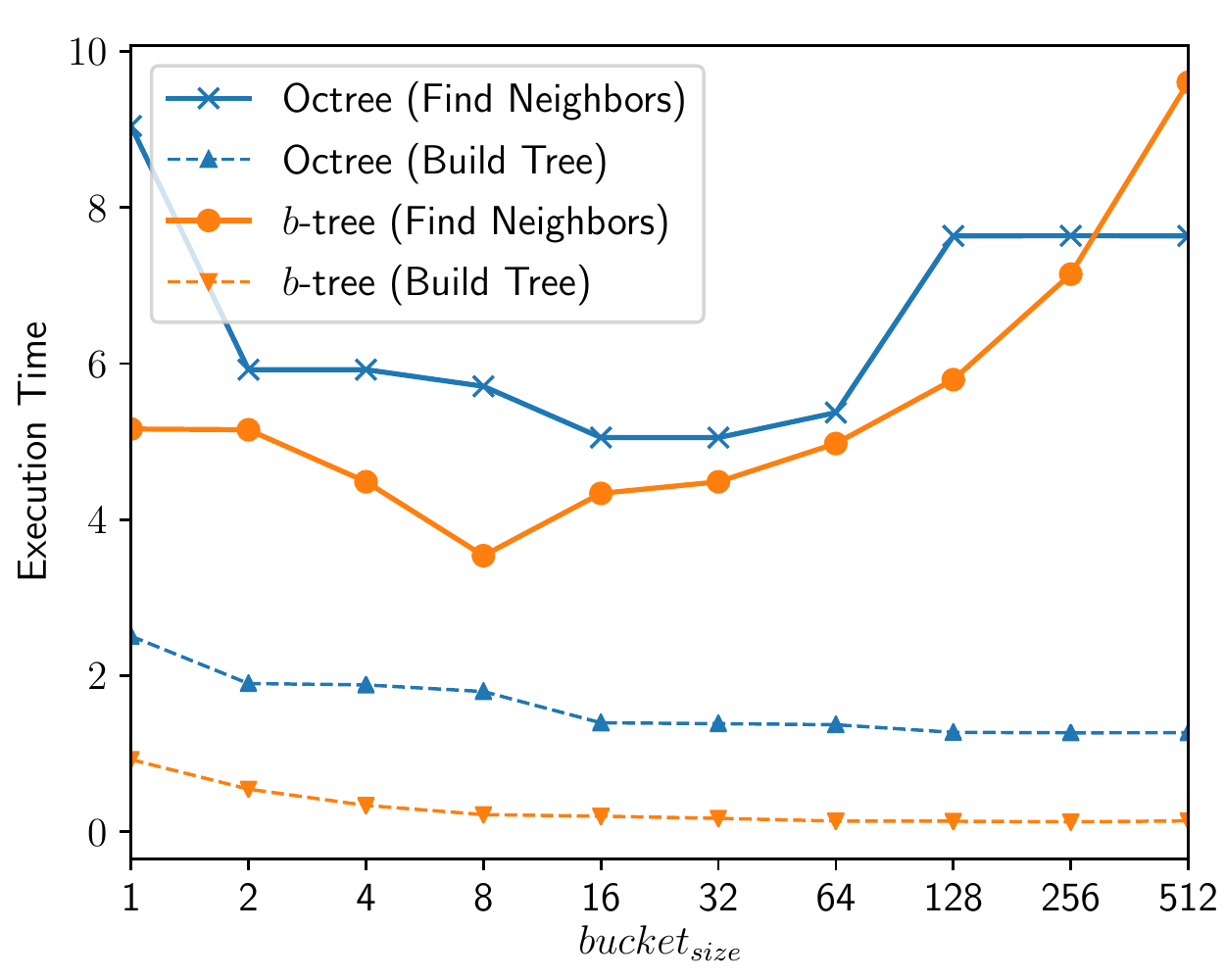}}
	\caption{Impact of the bucket size $S$ on the execution time of the tree-based algorithms for the EC test~(a) and the SP test~(b).}
	\label{fig.bucketsize}
\end{figure*}

In this sub-section, we evaluate the impact of the bucket size parameter $s$ on the performance of the \tree- and octree-based algorithms for the EC and SP datasets. 
The optimal value of the bucket size for each test is derived experimentally, as shown in Figure~\ref{fig.bucketsize}.
For building the tree, a larger bucket size means more particles per cell, and therefore less cells overall and a more shallow tree, which is faster to build.
For finding the neighbors, a small bucket size is preferred to avoid having too many particles per cell, which increases the number of particle to particle distance checks within a cell. The optimal value is around $8$ for both test cases.

\subsection{Impact of $\alpha$ and $\beta$}

\begin{figure*}[!htb]
	\centering
		\subfloat[Evrard Collapse]{\includegraphics[scale=0.5]{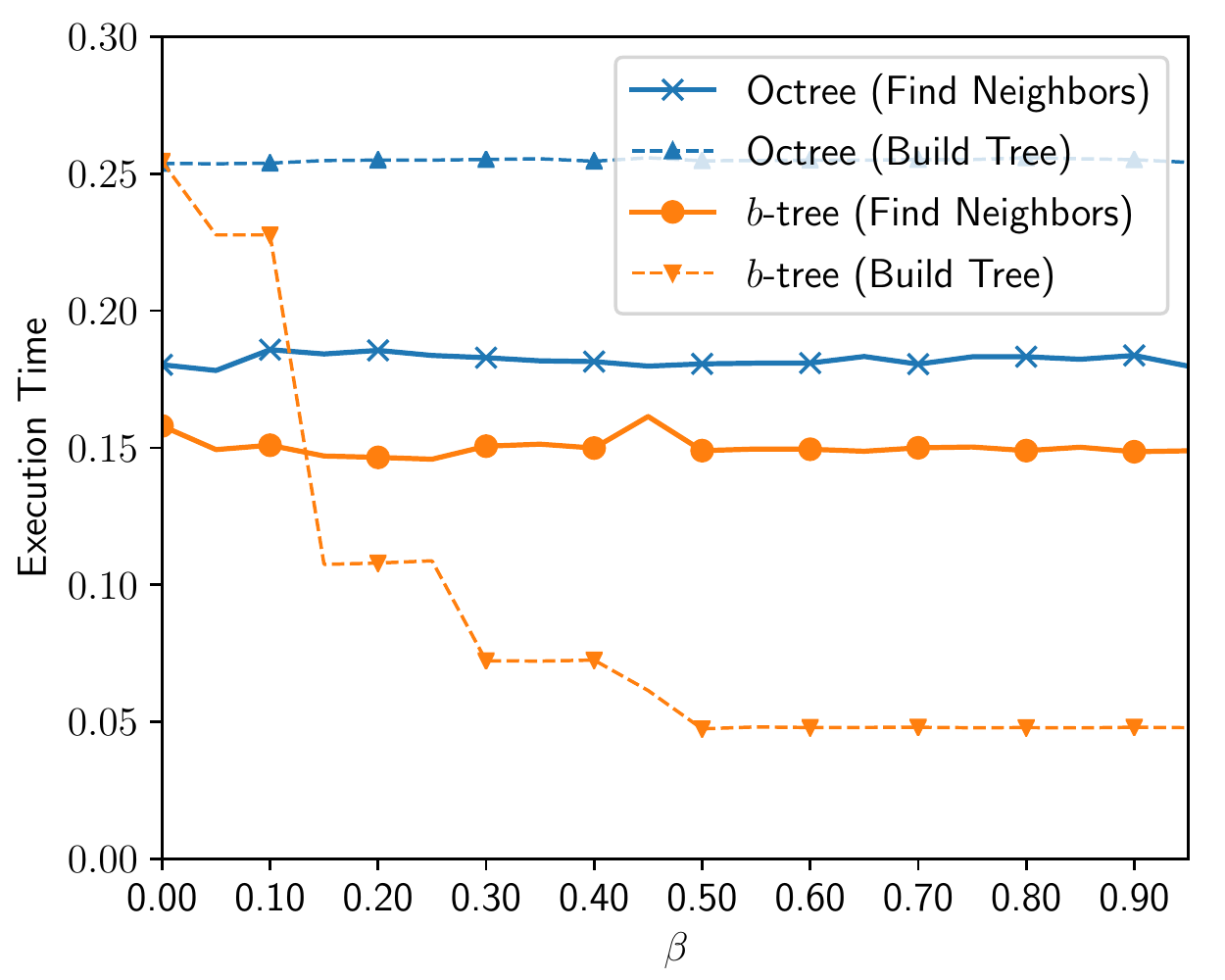}}
		\subfloat[Square Patch]{\includegraphics[scale=0.5]{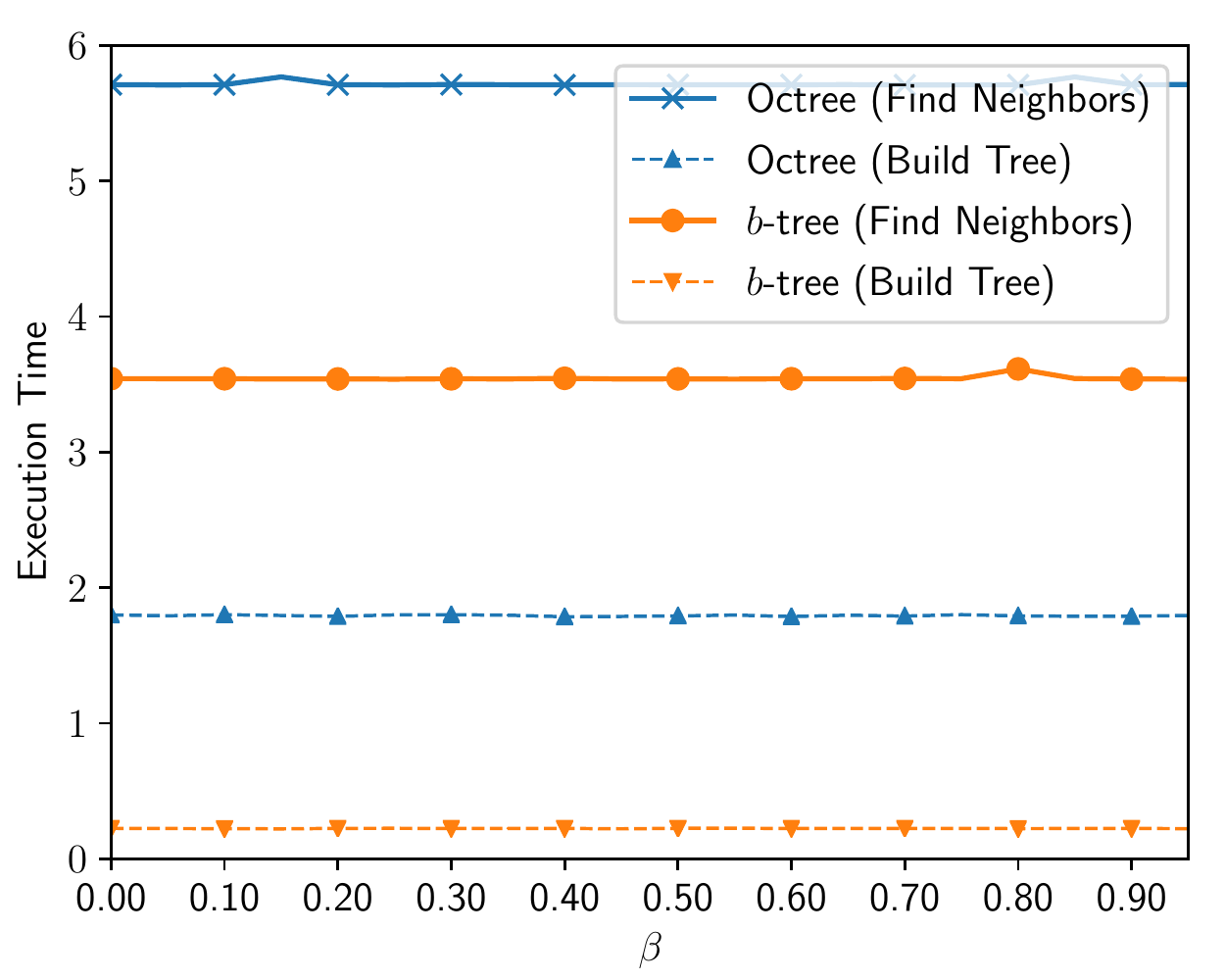}}
	\caption{Impact of the target ratio parameter $R$ on the execution time for Evrard Collapse test (a) and the Square Patch test (b).}
	\label{fig.ratio}
\end{figure*}

Finally, we investigate the impact of the $\beta$ parameter, which control the maximum amount of cells that have too few particles with respect to $\alpha\sim 0.5$, and hence we control the final branching factor for every tree node.
Figure~\ref{fig.ratio} shows both, the \find and \build execution time for different values of $\beta$.
The non-uniform particle distribution for EC test case means than many empty cells are created with the default branching factor. With small values of $\beta$, \build ends up selecting very small branching factors and the associated time is close to the classical octree behavior.
However with higher values for $\beta$, \build is allowed to use more cells and there are less redistributions steps, which is faster to build.

\subsection{Tree Walk Visualization}

\begin{figure*}[!htb]
	\centering
		\subfloat[\tree]{\includegraphics[scale=0.8]{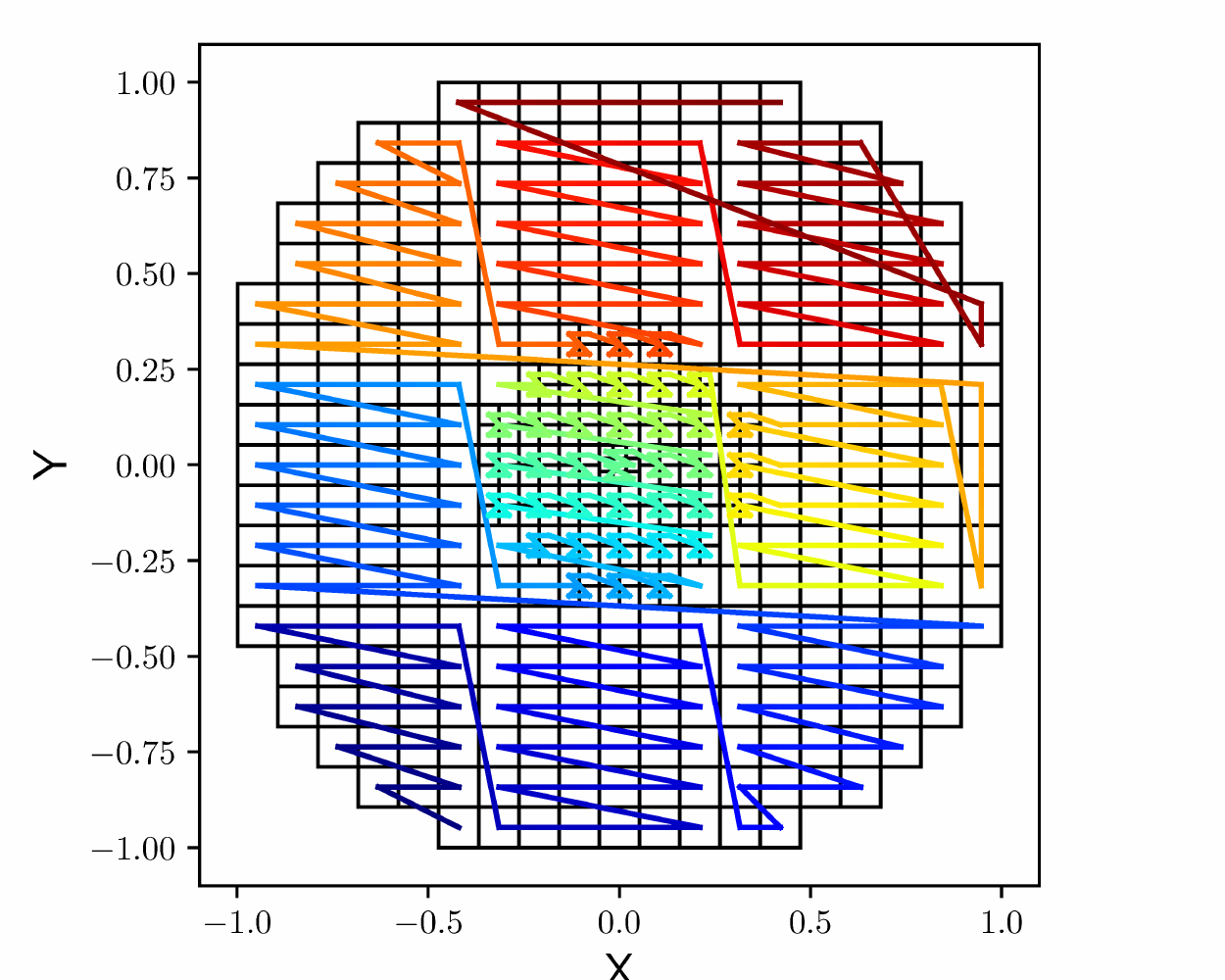}}\\
		\subfloat[octree]{\includegraphics[scale=0.8]{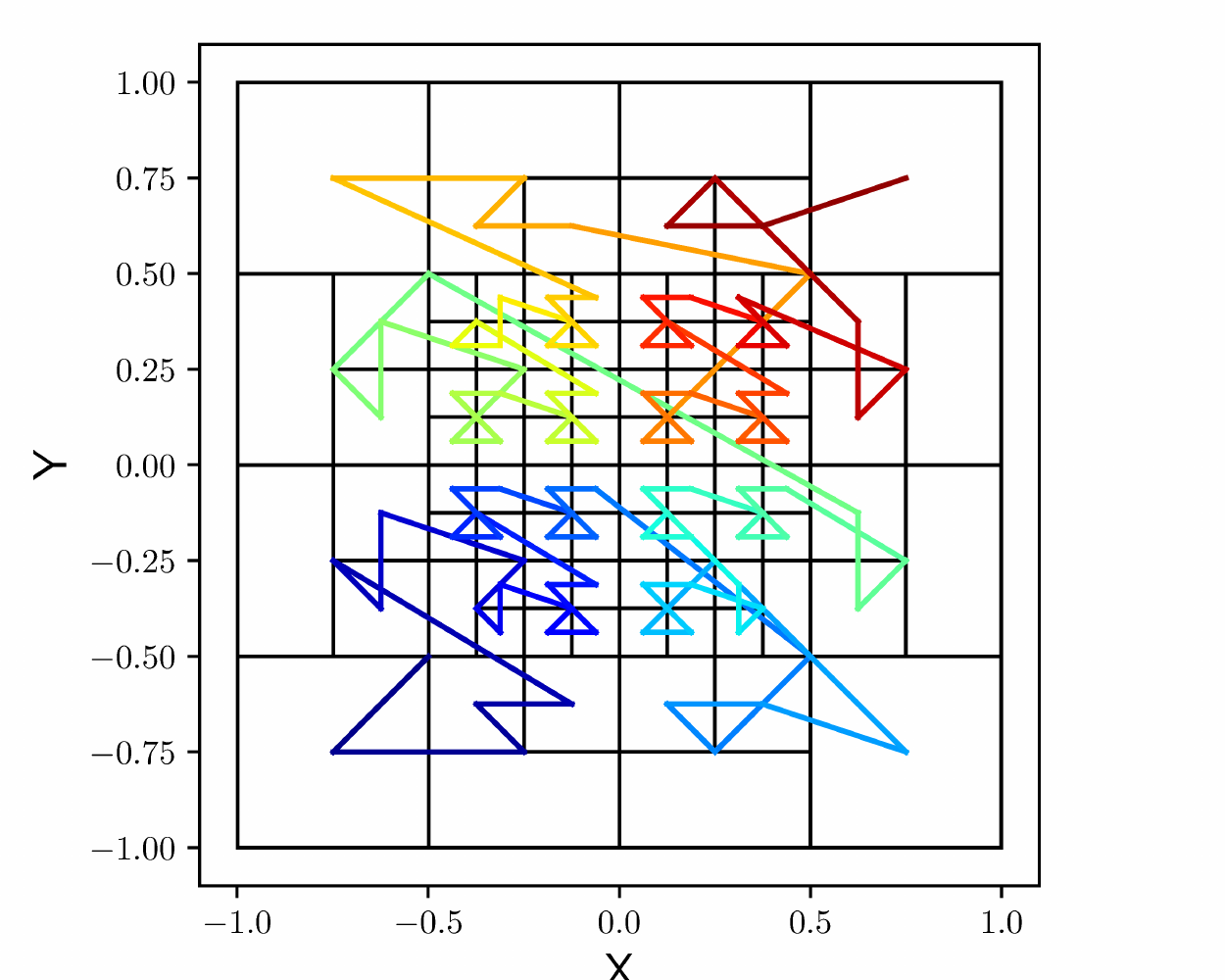}}
	\caption{2D slice at the center of the 3D domain for the EC test case. Black squares represent cells and the colored line shows the order in which the cells are visited when walking the tree. The bucket size is set to $s=32$ for more visibility, and particles are not shown.}
	\label{fig.line.evrard.octree}
\end{figure*}

Finally, we provide a visual comparison of the resulting trees for the non-uniform EC test case. Figure~\ref{fig.line.evrard.octree} shows a 2D slice of the 3D domain at the center.
Black squares represent cells and the colored line shows the order in which the cells are visited when walking the tree recursively.

Two important optimizations have been implemented to accelerate the tree walk: (1) particles are reordered in memory along this curve; and (2) \build and \find algorithms have been implemented with a blocked approach, similar to the one employed in blocked matrix-matrix multiplication implementations in order to further improve cache reuse. Different colors correspond to different blocks.
Note that while \tree has only two levels (note the two different cell sizes), the octree has $5$ levels.



\subsection{Performance Optimizations} 
This sub-section presents a performance analysis for the \find algorithm for \tree and octree comparing them in terms
of execution time and degree of load imbalance. The experiments employ five OpenMP loop scheduling
strategies proposed and described in recent work by~\cite{Ciorbaschedulers}:
\texttt{static}, \texttt{dynamic}, \texttt{guided}, \texttt{fac2} and
\texttt{rand}.

\begin{figure}[!htb]
	\centering
		\includegraphics[scale=0.7]{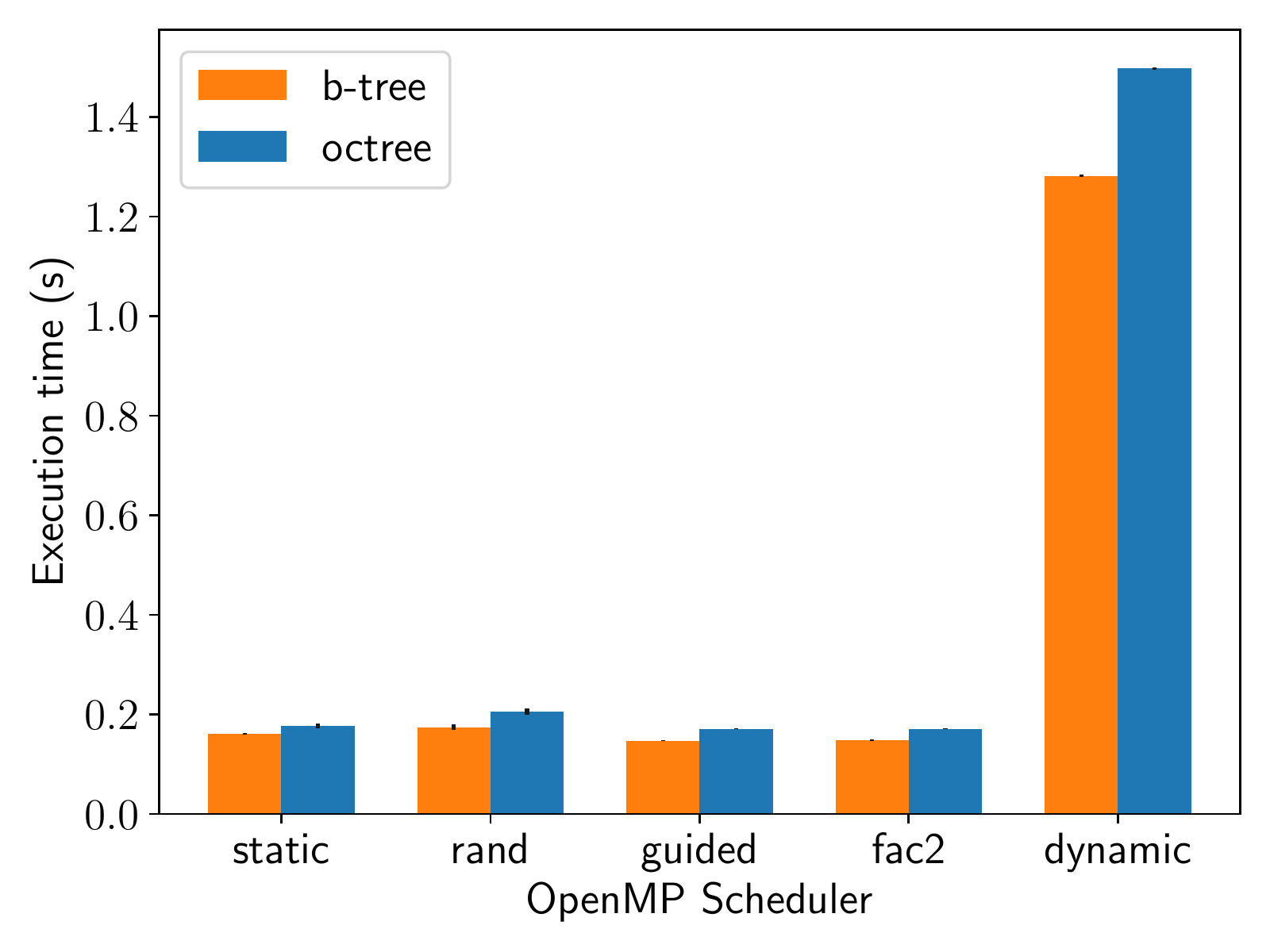}
	\caption{Execution time of employing \tree and octree for the EC test case.}
	\label{fig.exectime.cov.evrard}
\end{figure}

\begin{figure}[!htb]
	\centering
		\includegraphics[scale=0.7]{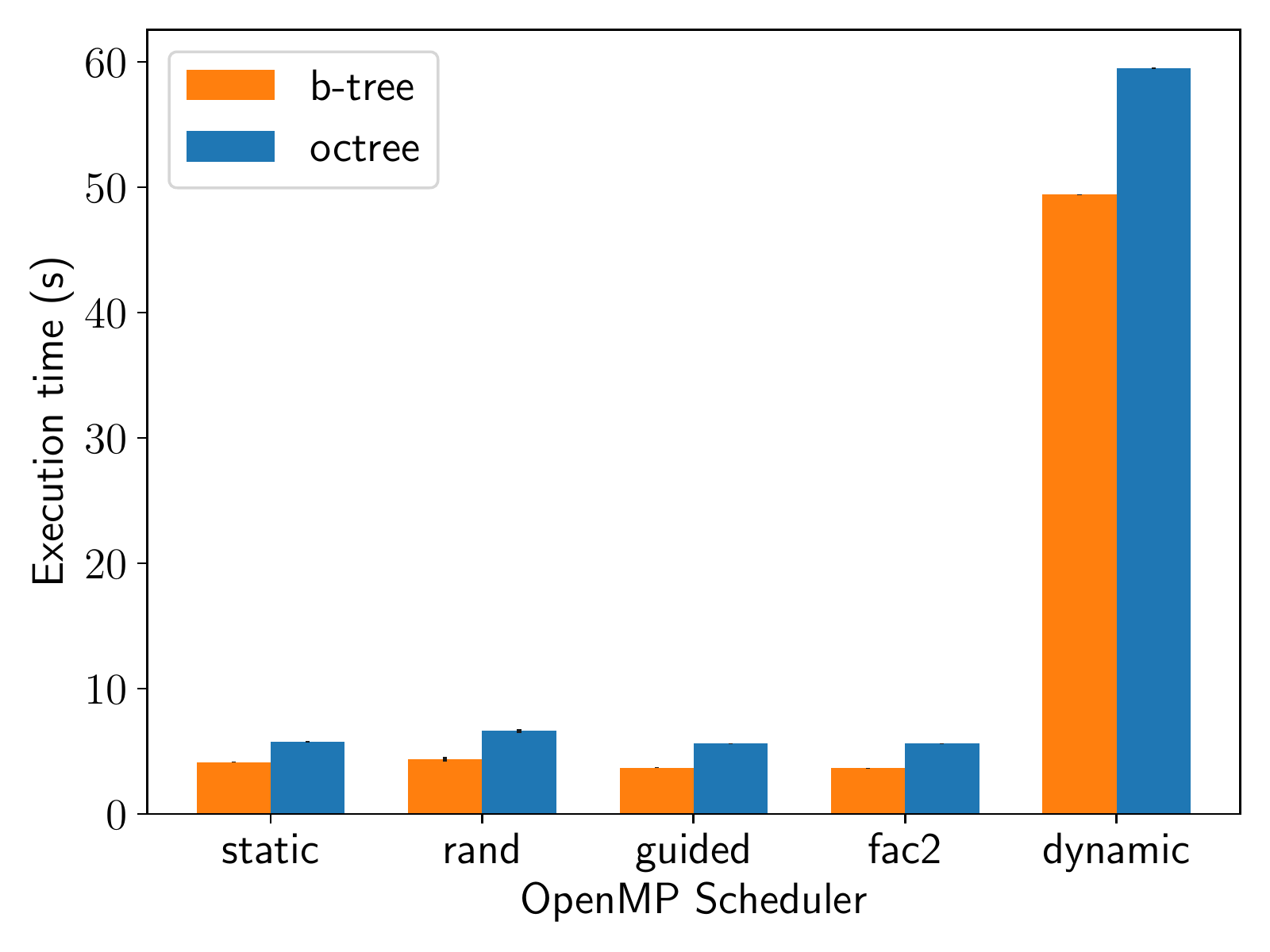}
	\caption{Execution time of employing \tree and octree for the SP test case.}
	\label{fig.exectime.cov.squarepatch}
\end{figure}




Figure~\ref{fig.exectime.cov.evrard} presents the average execution time or the EC input. 
In terms of execution time, \tree obtained better results
for all scheduling strategies. The later approaches an execution time of 0.15
seconds using most scheduling strategies except for \texttt{dynamic}, when it
reaches $1.3$ seconds. In terms of load balancing, \texttt{dynamic} always
balances the load better, while causing significant overhead that noticeably
degrades the performance.

Figure~\ref{fig.exectime.cov.squarepatch} presents the average execution
time the results for the SP test case.
In Figure~\ref{fig.exectime.cov.squarepatch}, 
\tree obtained better results for all scheduling
strategies. However since this input is much larger, the execution time stays
around 4 seconds for most of the scheduling strategies, except for
\texttt{dynamic} that causes significant overhead during the execution. 
The small degree of load imbalance observed is well amortized by the
benefit of the self-scheduling properties underlying \texttt{guided} and
\texttt{fac2}, resulting in improved performance.

\subsection{SPHYNX}
\label{sec.sphynx}

\begin{figure}[!htb]
	\centering
		\subfloat{\includegraphics[scale=0.8]{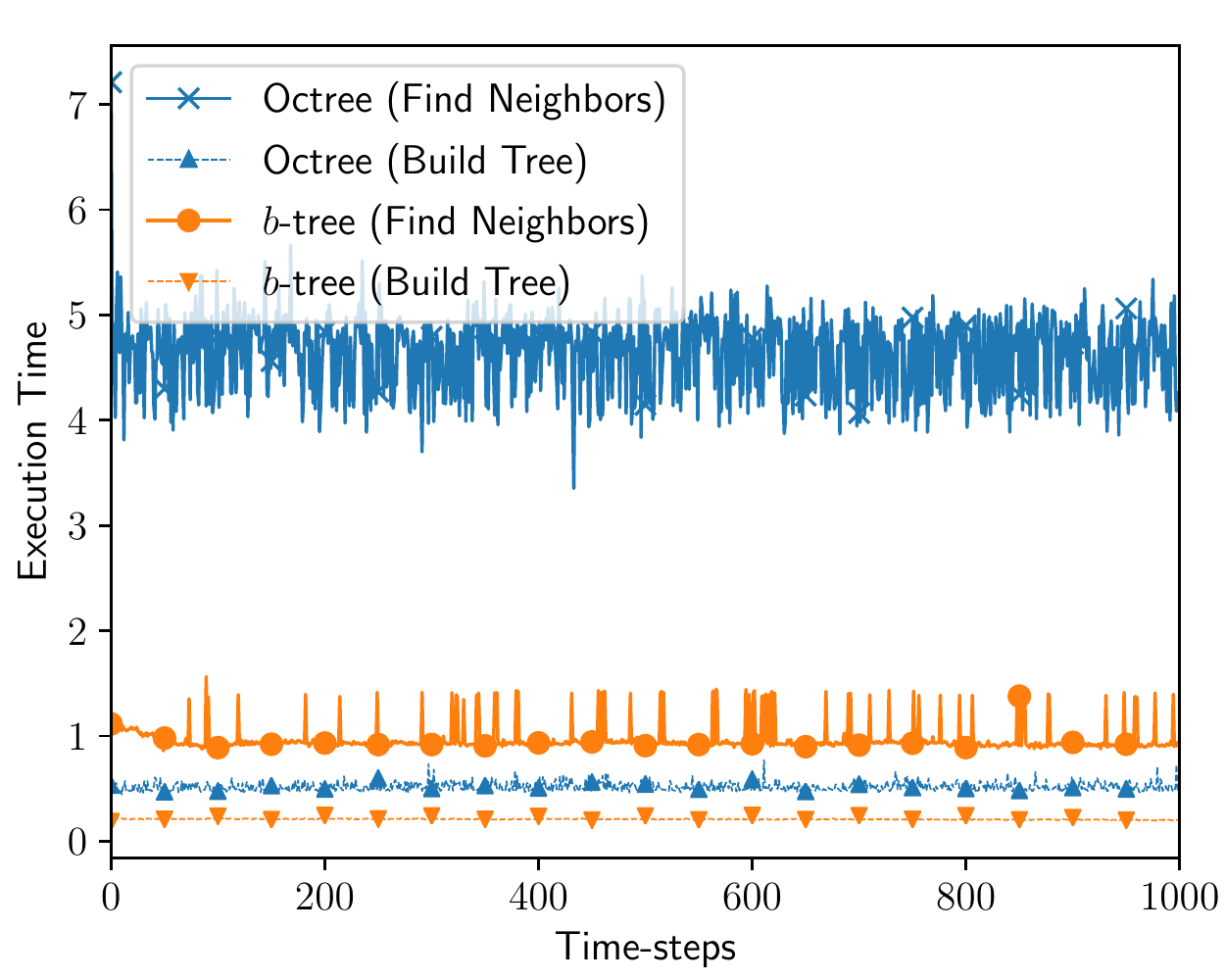}}
	\caption{SPHYNX execution times for \find and \build algorithms compared to the legacy octree implementation.}
	\label{fig.sphynx}
\end{figure}

Finally, \tree has been integrated into an astrophysical SPH code, SPHYNX~\cite{cabezon2017}. Figure~\ref{fig.sphynx} shows the execution time of the \tree algorithms with respect to the original, legacy octree implementation for a smaller SP test case with one million particle (the 10 million particles test could not run with the legacy code). \find is up to $5\times$ faster.



\section{Conclusion}
\label{sec:conclusions}

In this paper, a novel tree structure, namely \tree, was proposed to improve the exact close neighbors searching time for Smoothed Particle Hydrodynamics simulations. Algorithms to build the tree and to find the neighbors have been presented, and their complexity analyzed.
Experimental results show both a good scalability and improved speedup compared to a classical octree implementation.
Integration of the algorithm in a production SPH simulation shows promising results, with up to $5\times$ improvements over the legacy code, delivering relevant speedups in a section of the hydrodynamical codes that is a common performance bottleneck.

Future work will address computational domain decomposition among computing nodes, coupling of the tree-based algorithms with gravity calculations, as well as testing the usefulness of the proposed \tree on other SPH test cases.

\section*{Acknowledgement}
\label{sec:ack}
This work is supported in part by the Swiss Platform for Advanced Scientific Computing (PASC) via the ``\mbox{SPH-EXA}: Optimizing Smooth Particle Hydrodynamics for Exascale Computing'' grant and by the Swiss National Science Foundation (SNF) via the ``Multi-level Scheduling in Large Scale High Performance Computers'' grant, number 169123. The authors acknowledge the support of sciCORE (http://scicore.unibas.ch/) scientific computing core facility at University of Basel, where part of these calculations were performed. 

\bibliographystyle{abbrv}
\bibliography{biblio}

\end{document}